\documentstyle[aps,amssymb,epsfig]{revtex}

\begin{document}

\title{Non-linear harmonic generation in finite amplitude
black hole oscillations}

\author{Philippos Papadopoulos}

\address{School of Computer Science and Mathematics,
University of Portsmouth,
PO1 2EG, Portsmouth, United Kingdom}
\maketitle

\begin{abstract}
The non-linear generation of harmonics in gravitational perturbations
of black holes is explored using numerical relativity based on an
in-going light-cone framework. Localised, finite, perturbations of an
isolated black hole are parametrised by amplitude and angular harmonic
form. The response of the black hole spacetime is monitored and its
harmonic content analysed to identify the strength of the non-linear
generation of harmonics as a function of the initial data
amplitude. It is found that overwhelmingly the black hole responds at
the harmonic mode perturbed, even for spacetimes with 10\% of the
black hole mass radiated. The relative efficiencies of down and
up-scattering in harmonic space are computed for a range of
couplings. Down-scattering, leading to smoothing out of angular
structure is found to be equally or more efficient than the
up-scatterings that would lead to increased rippling. The details of
this non-linear balance may form the quantitative mechanism by which
black holes avoid fission even for arbitrary strong distortions.
\end{abstract}


\section{Introduction}
\label{intro}

The dynamical behaviour of black holes near equilibrium has been
clarified using classic, technically elaborate, perturbation theory
(see e.g.,~\cite{Chandra}). In recent years there has been an interest
in studying more general aspects of the dynamics of black holes as
they would manifest e.g., in the merger of two black
holes~\cite{bbhgc}, or in other large deformation induced by an
external agent. In this regime there are no readily available analytic
tools and modern approaches advocate the use of discretization
procedures. The most explored computational framework for the study of
black hole dynamics dynamics~\cite{smarr} is facing a number of
obstacles which limit the duration and accuracy of black hole
simulations. Irrespective of difficulties, persistent work within this
approach in the past decade~\cite{abrahams,anninos,bernstein} has
uncovered important aspects of finite amplitude dynamics of black
holes. A central suggestion of this effort has been that, in a larger
than expected part of the parameter space, the physics of finite
perturbations of black holes can be expressed in the language of
infinitesimal perturbation analysis. It will be useful for what
follows to review the main elements of this language: Linearised black
hole perturbations are mapped into a problem of scattering off a
positive potential which results from a combination of the angular
momentum barrier with the one-way absorbing membrane of the
horizon. Important role in the perturbative drama play the
exponentially damped oscillating modes called quasi-normal modes
(QNM). Those solutions form the mechanism by which {\em weak}
perturbations of a black hole are radiated away, leading to a
stationary remnant~\cite{price}. 

The discovery in~\cite{abrahams,anninos,bernstein} that finite
perturbations of black holes, seen in either black hole-black hole or
black-hole-gravitational wave systems, seem to emit their energy
primarily through a linear channel was subsequently illuminated
further with the interpretation of the numerically generated black
hole spacetimes from the point of view of perturbation
theory~\cite{price,gleiser}. There have been a number of more general
studies~\cite{brandt,camarda} investigating the validity of the basic
picture for rotating and three-dimensional perturbations of black
holes. This pioneering numerical work helped sharpen a number of
questions surrounding non-linear black hole dynamics. Is there a
genuinely non-linear regime? Is it visible to remote observers? What
are the salient features of the evolution of an arbitrarily distorted
black hole? Most intriguingly, how exactly does the non-linear
evolution of a highly distorted black hole manage to avoid the perils
of a fission and hence conform with the area theorem
expectation~\cite{hawking}?

The present work aims to tackle those questions by approaching the
computational challenge from a different angle, namely by using a
geometric approach based on the characteristic initial value problem
(civp). The civp was introduced in seminal work by Bondi and
Sachs~\cite{bondi,sachs} as an asymptotic, but non-perturbative,
analysis of radiating spacetimes. Several variations have been
proposed (see e.g.,~\cite{TW,dinv,bartnick}) to enable global
computations of general spacetimes. Those formulations typically share
an excellent economy and adaptability to computations of
quasi-spherical radiative spacetimes. Versions of the civp have been
extensively used for the study of black hole dynamics in spherical
symmetry, in connection with matter fields in particular (see. e.g.,
~\cite{gundlach,stewart,pf1}). Higher dimensional studies based on the
civp have been rapidly maturing over the past decade. An easily
accessible review is given in~\cite{living}.

The work presented here is based on the formal framework described in
~\cite{TW}. The first adaptation of this formulation to the study of
black holes using ingoing light-cone foliations was presented
in~\cite{marsa}, in a spherically symmetric setting.  Subsequent
developments led to long term stable three-dimensional computations of
black holes~\cite{stable3d}. In this paper, an algorithm developed
originally for the study of regular axisymmetric spacetimes~\cite{JMP}
has been adapted and applied to the study of finite amplitude black
hole perturbations. The approach is described in Sec.~\ref{method}.
The details of the numerical algorithm are substantially the same as
in~\cite{JMP} and will not be repeated. The emphasis of the discussion
will be instead on the new elements which adapt the method to the
problem at hand. This includes, firstly, initial and boundary
conditions and, secondly, a method for extracting the relevant
physical information. Given the substantially different nature of
solutions explored here compared to the tests in~\cite{JMP},
re-calibrations and accuracy tests have been performed and pointed out
where needed. In Sec.~\ref{results} the results of a parameter space
survey covering initial data amplitude and a portion of the harmonic
space are presented. The discussion begins with the study of selected
waveforms and leads to the main topic, which is the quantification of
energy transfered by various non-linear couplings. The relevance of
the findings to the questions that motivated this work is
 assessed in the conclusions~\ref{summary}.

The usual unit conventions $(G=c=1)$ apply. The spacetime signature
has been modified from timelike~\cite{JMP} to spacelike.

\section{Geometric and computational setup}
\label{method}

\subsubsection{Framework}

\begin{figure}[t]
\vspace{0.5cm}
\centerline{\epsfig{figure=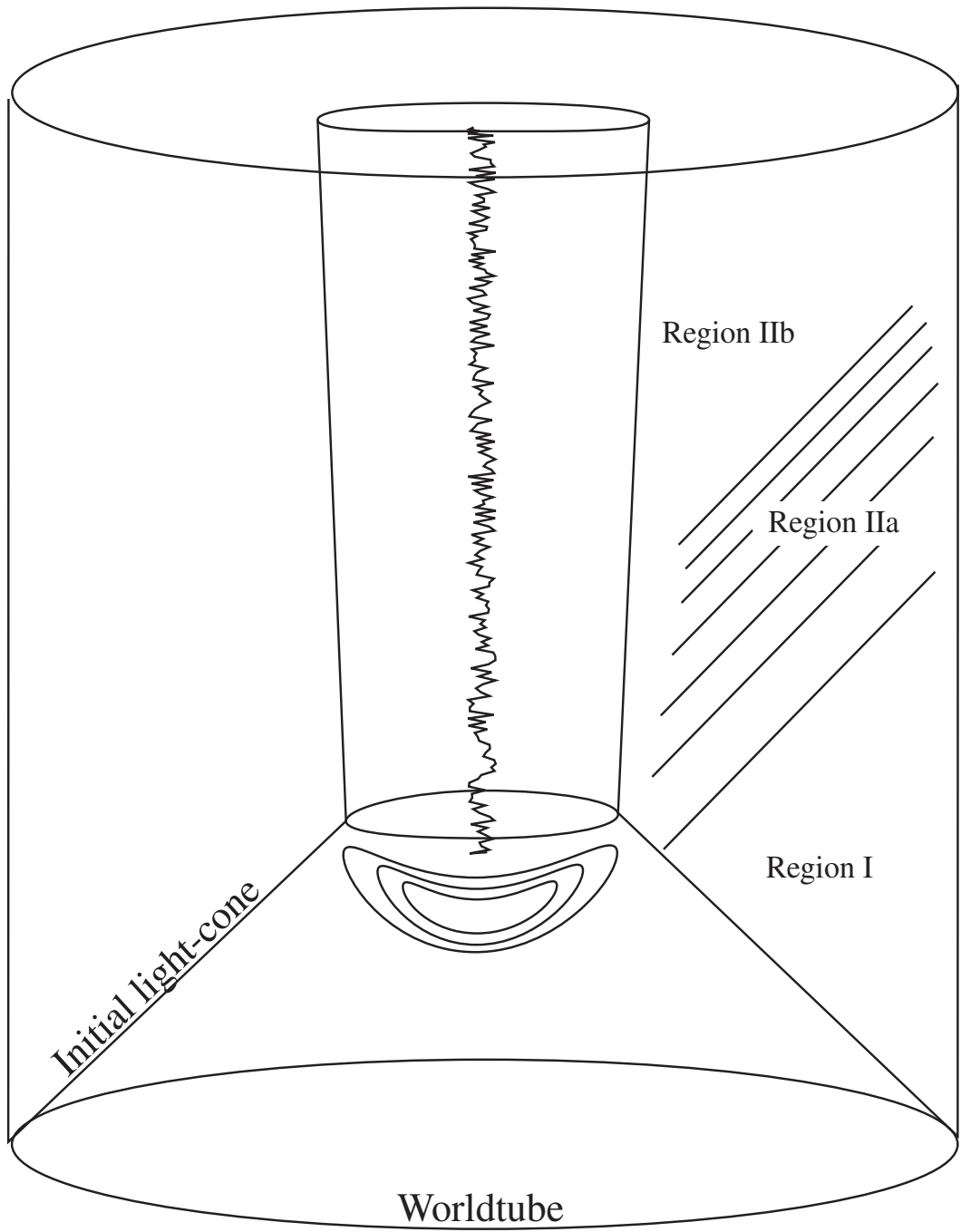,height=18cm,width=14cm}}
\vspace{0.5cm}
\caption{Spacetime diagram illustrating the main aspects of the
geometric and computational setup of the problem. The foliation is
based on advanced time, using ingoing light-cones which emanate from a
world-tube. The geometry at the world-tube is kept fixed at all times
and is given by the Schwartzschild values. The initial ingoing light
cone is distorted with the specification of an arbitrary amount of
shear. The sole constraint comes from the requirement that the ingoing
cone will not develop a caustic before one reaches a marginally
trapped surface. The evolution generates a dynamic spacetime, with the
initial data scattering both towards and away from the black hole. For
initial data that are restricted to zero outside some given radius,
there is clean propagation into the unperturbed region I, where the
spacetime is Schwartzschild. The outgoing gravitational disturbance
(region IIa) can be intercepted by an ``observer'' at a suitable
radius and analysed with respect to structure. At late times, of the
order of hundreds of M, a slowly evolving ``remnant'' appears in the
inner region IIb. The portion of the spacetime that emerges there
represents a {\em distorted} black hole, the agent of distortion being
the outward propagating gravitational radiation energy that intervenes
between the horizon and the spherical worldtube.}
\label{Diagram}
\end{figure}

The algorithm is based upon the civp for the Einstein equations in vacuum,
using light cones emanating from a timelike worldtube.  With the
conventions of the Bondi-Sachs gauge, the explicit form of the
metric element is
\begin{equation}
 ds^2 =  - ((1 - {2 M \over r})  e ^{2 \beta} - U^2 r^2 e^{2 \gamma}) dv^2
           - 2 e^{2 \beta} dv dr
           + \  2 U r^2 e^{2 \gamma} dv d\theta \nonumber 
       + r^2 (e^{2 \gamma} d\theta^2 + e^{-2\gamma} \sin^2 \theta d\phi^2).
\end{equation}
This form implies an axisymmetric spacetime with zero rotation. The
metric variables $(\gamma,M,\beta,U)$ are functions of the coordinates
$(v,r,\theta)$ only. The formulation of a boundary initial value
problem follows the lines of~\cite{TW}, with light-cones 
emanating from a timelike worldtube at a finite radius and proceeding
inwards, to intercept the black hole horizon (see
Fig.~\ref{Diagram}). This setup may be considered as a straightforward
generalisation of the well known family of ingoing
Eddington-Finkelstein coordinate system to axisymmetric distorted
black holes.

The vacuum Einstein equations decompose into three hypersurface
equations and one evolution equation (the conservation equations along
the world-tube will not be used and are omitted here). The overall
form of the equations is given in symbolic notation (the reader is
referred to~\cite{IWW,JMP} for the detailed expressions as they are
used in the code):
\begin{eqnarray}
\label{evolve}
 \Box\, ^{(2)} \psi & = &
{\cal H_{\gamma}}(M,\beta,U,\gamma), \\ \label{h1}
 \beta_{,r} &=& {\cal
H}_\beta(\gamma), \\ \label{h2}
 U_{,rr} & = & {\cal H}_U(\beta,\gamma) , \\ \label{h3}
 M_{,r} & = &{\cal H}_M(U,\beta,\gamma) ,
\end{eqnarray}
where $\Box\,^{(2)}$ is an appropriate 2-dimensional wave operator
acting on the $(v,r)$ sub-manifold, $\psi=r\gamma$, and the symbols
$H_{X}$ denote reasonably lengthy right-hand-side contributions. Free
initial data for $\gamma$ on an ingoing light-cone $\cal{N}$ lead,
with the integration of the radial hypersurface
equations~(\ref{h1},\ref{h2},\ref{h3}), to the complete metric along $\cal{N}$.
This then enables the computation of $\gamma$ on the next light cone,
with the use of the evolution equation~(\ref{evolve}). This system of
equations is extremely economical, given that it encodes the full
content of the Einstein equations for the assumed symmetry.

\subsubsection{Linearised limit}

Before proceeding to the details of setting up the initial value
problem, there is some merit in discussing the linearised limit of the
equations~(\ref{evolve},\ref{h1},\ref{h2},\ref{h3}) around a spherical black hole
background of mass $M_{0}$. The linearised limit around Minkowski
spacetime ($M_{0}$) was analysed in~\cite{bondi,JMP} and various
reductions of the dynamics to the wave equation have obtained. For
non-zero background mass, one obtains the coupled system
\begin{eqnarray}
\label{linear}
2 \psi_{,rv} + ((1-\frac{2M_0}{r}) \psi_{,r})_{,r} & = &
- \frac{2M_{0}}{r^2} \gamma - \frac{1}{2r} (r^2 \sin\theta 
(\frac{U}{\sin\theta})_{,\theta} )_{,r} \, ,\\
(r^4 U_{,r})_{,r} & = & - 2 \frac{r^2}{\sin^2\theta}
(\sin^2\theta \gamma)_{,r\theta} \, .
\end{eqnarray}
Physically, from the assumptions leading to the metric element, those
equations describe axisymmetric, {\em even parity} perturbations of
Schwartzschild black holes~\cite{Chandra}. They also ought to be
reducible to a single equation in the spirit of perturbation theory.
The presence of a black hole affects the evolution of the shear of the
ingoing light-cone primarily with the modification of the slope of
outgoing characteristics, which vanishes at the horizon. There is a
potential term, specific to the spin-2 gravitational
perturbations. The remaining RHS term in the $\gamma$ equation,
together with the $U$ equation, form effectively the angular momentum
barrier.

With the minimal structure of the equations now transparent, in the
restoration of non linear terms, one can identify at least three
distinct alterations: Firstly, a modification of the light-cone
structure, affecting the principal part of the equation and hence the
propagation speed of signals. Secondly, a modulation of the existing
spherical potential term. For example, the mass aspect function $M$
acquires angular structure. Thirdly, quadratic couplings emerge and
act as source terms, e.g., in the evolution equation. The combined
effect of the last two agents will be the mixing of oscillation modes
of different geometrical characteristics.

\subsubsection{Boundary and Initial Data}

Within an ingoing light-cone framework, there are two possible
prescriptions for setting up initial and boundary data for the study
of the response of a black hole to perturbations. In the first, akin
to a {\em scattering} study, the initial advanced time light cone
$\cal{N}$ has trivial background data. Incoming gravitational
radiation is introduced at the world tube $\cal{W}$ through the
specification of the unconstrained $\gamma(v,\theta)|_{\cal{W}}$
function. A self-consistent evolution of the boundary data is then
achieved by the time-integration of the conservation conditions along
the world-tube. In the second approach, which may be called a {\em
perturbation} study, the free data are specified on $\cal{N}$, whereas
$\cal{W}$ is kept with trivial data. The two approaches are closely
linked, since the scattering of incoming radiation will immediately
lead to outgoing perturbations. Still, the perturbation approach is
computationally simpler, since it avoids the integration of the
conservation conditions. This is the approach used here.

The world-tube boundary conditions are simple but central to the setup
of the physical problem. A non-rotating unit-mass black hole is
prescribed at $\cal{W}$ by setting $M(v,r_{\cal{W}})=1$ and all other
fields equal to zero. Such a condition is not in general compatible
with a spacetime in which outgoing radiation is filtering through the
worldtube. Minimally, such radiation would be diminishing the mass
contained inside the worldtube, and $M(v,r_{\cal{W}})$ would not be
constant. Hence, for consistency, evolutions must be restricted in
time so that the neighbourhood of the world-tube is unperturbed by
local flux of radiation. Fulfilling this condition is entirely
possible and will always be the case in the results presented here.

The free initial data on the ingoing light cone $\cal{N}$ are captured
by a single real function of two variables $\gamma(r,\theta)$. For a
spherical unperturbed spacetime this function is identically zero. The
radial profile of a perturbation must be chosen so that the evolution
conforms with the constraint discussed in the previous paragraph. This
leads to the adoption of an exponentially decaying profile, as a model
of a localised disturbance superposed on a black hole. Clearly, for
perturbations arising in astrophysical systems, the prior history of
the system is likely to have produced a more extended perturbation,
arguably though, of very small amplitude. A Gaussian profile, centred
at a radius $r_c$ is hence chosen as a generic representative, and this
freedom will not be explored here any further. Comparison between
different values for $r_{c}$ suggests that values close to $r=3M$ are
more effective in generating outgoing non-linear response and this
value will be adopted throughout.

The {\em angular profile} of the initial data will play a key role in
the following discussion. The only constraint is that the function a
has suitable falloff at the pole $y=\pm1$, consistent with the fact
that $\gamma$ is a spin-2 scalar representing the everywhere regular
intrinsic metric of topological spheres of constant r. This fall-off
condition is equivalent to requiring that the angular decomposition of
$\gamma$ in harmonics starts with at least a quadrupole term. The spin
harmonics relevant to this decomposition are discussed in detail
in~\cite{Penrose}. In axisymmetry, a convenient expression for the
basis functions is given by
\begin{equation}
Y_{2l0}= \left(\frac{(l-2)!(2l+1)}{4\pi(l+2)!}\right)^{1/2} 
(P_{l,\theta\theta} - \cot\theta P_{l,\theta})
\end{equation}
where $Y_{2lm}$ denotes an $(l,m)$ harmonic of spin $2$ and $P_{l}$
are the Legendre polynomials.

The free specification of initial data, one of the virtues of the
civp, permits the introduction of {\em pure} angular harmonic
dependence, by means of expressions like, e.g.,
\begin{equation}
\gamma(r,\theta) = \frac{\lambda}{\sqrt{2\pi}\sigma} e^{- (r - r_c)^2/\sigma^2} Y_{2l0}(\theta) \, .
\label{indata}
\end{equation} 
The multipole index $l$ of the initial data will be referred to as the
{\em primary harmonic}, as it is to be expected that it will play an
important role in the response of the black hole geometry to the
perturbations. The physical content of such data is essentially a
local distortion of the ingoing light-cone, with prescribed angular
structure and amplitude. The claim that the spacetime constructed by
this approach is actually a distorted black hole is made geometrically
more precise by the introduction of the concept of a marginally
trapped 2-surface (MTS) on a given ingoing light-cone $\cal{N}$. An
MTS is defined, in this context, as the two-parameter radial function
$R(r,\theta)$ on which the expansion
$\Theta=2\nabla^{\alpha}l_{\alpha}$ of an outgoing null ray pencil
$l^{\alpha}$ vanishes~\cite{marsa}. Using a relaxation method for the
solution of equation $\Theta=0$ one obtains indeed a distorted initial
MTS, showing curvature dependent on the amplitude of the data. The
invariant analysis of the intrinsic geometry and dynamics of the MTS
are interesting problems in themselves and deserve separate study.

\subsubsection{Extraction of the non-linear response}

The evolution of the initial data leads, in general, to radiation
emission towards both infinity and the horizon. Here the focus will be
on the outgoing, in principle observable, part. The non-linear
response of the black hole to the initial perturbation will be encoded
quite legibly in the angular harmonic decomposition of the outgoing
solution. A strategy to isolate this information is outlined here.

The metric perturbation encoded in $\gamma$ can be decomposed in
spin-2 harmonics as
\begin{equation}
\gamma(v,r,\theta) = \sum^{\infty}_{l=2} \gamma_{l}(v,r) Y_{2l0}(\theta) \, , 
\end{equation}
which is inverted as 
\begin{equation}
\gamma_{l}(v,r) = 2 \pi \int^{1}_{-1} \gamma(v,r,y) Y_{2l0}(y) dy \, ,
\end{equation}
using the orthonormality of the spin-2 harmonics and introducing
$y=-\cos\theta$. At this stage the decomposition is non-perturbative,
i.e., it does not depend on $\gamma$ being small and can be can be
carried out across the entire light-cone. The set of functions
$\gamma_{l}(v,r)$ is effectively capturing {\em all} the information
about the given spacetime.  For non-linear evolutions, this set
includes more than one harmonics, even if the initial data are set to
a single (the primary) harmonic dependence. The new harmonics will be
referred to as the {\em secondary harmonics}. In numerical practice,
this decomposition is obtained via an eighth-order accurate (in
$\Delta y$) angular integration of the grid data.

A good indicator of non-linear processes near the black hole is the
amount of energy emitted in secondary harmonics. In the present setup
the foliation does not extend to infinity, hence the extraction of an
energy estimate for an outgoing solution must resort to an approximate
expression. The task is straightforward, due to the adapted form of
the metric element (i.e., spherical coordinates, light-cone gauge).
Further, sufficiently far from the black hole (but inside the
world-tube), the outgoing radiation is weak in amplitude and the
spacetime is close to a Schwartzschild spacetime in its standard
ingoing Eddington-Finkelstein form. At that radius, denoted as the
observer radius $r_{o}$, and forming a timelike worldtube of area $4
\pi r_{o}^2$, the {\em news} can be approximated by $c(v) = \psi(v) =
r_{o} \gamma(v) $. From this, it follows that the total (negative)
energy flux crossing the observer will be approximating Bondi's energy
flux integral
\begin{equation}
\frac{dE_{tot}}{dv} = - \frac{1}{4\pi} \oint (\psi_{,v})^2
\sin\theta d\theta d\phi \, . 
\end{equation}
An expression for the total energy emitted in each angular mode $l$
after an evolution to a final time $T$ is given by
\begin{equation}
E_{l}(T) =  \frac{1}{4\pi} r^2_{o} \int^{T}_{0} (\gamma_{l,v})^2 dv \, .
\end{equation}
Given the availability of the complete spacetime, the device developed
here is but a partial probe whose usefulness rests on its quantitative
nature.

Other non-linear effects will include, for example, deviations of the
oscillation frequencies from the weak field limit. Those effects need
careful differentiation from the natural inclusion of higher
overtones, which are present in the initial response of linearised
perturbations as well. They also appear, at first sight, difficult to
quantify and for this reason will play a lesser role in this study.

\subsubsection{Computational aspects}

The computational algorithm follows closely that developed
in~\cite{JMP} and differs effectively only in the boundary treatment
and in the sense in which the integration proceeds (here towards
smaller radii). The algorithm was shown second order accurate in the
non-linear regime using a static boost-symmetric solution (SIMPLE) and
in the linear regime using exact decaying multipole solutions for
harmonics up to $l=6$. In addition the consistency of the evolution
was checked using global energy conservation.  The black hole dynamics
problem is sufficiently different and computationally
challenging~\cite{amr} though, that a re-calibration of the code has
been performed. It was found that, for the high resolution grid used
below, the primary harmonic can safely be captured to better than
1\%. The good overall accuracy of the code does not automatically
guarantee the correct capturing of the non-linear effects, as those are
{\em emergent}. For this reason, quantitative statements about such
effects will be individually certified as will be shown below.

The solution procedure involves about 350 lines of essential code. The
numerical grid is equidistant in both the radial $r$ and the angular
$y=-\cos\theta$ coordinates. Accurate calculations involve 2000 radial
grid points covering the radial domain from $r_{\mbox{min}}=1.8M$ to
$r_{\mbox{max}}=60M$ and 90 grid points covering the angular domain
$y=[-1,1]$. This corresponds to a $\Delta r = 0.03M$ and $\Delta
\theta = 0.02$ at the equator.  Evolution for a total time of $120M$
consumes about 30min of a 666Mhz Alpha CPU, at roughly 240
Mflops. About 45\% of that time is spend on the update of the
dynamical variable $\gamma$, with the rest being distributed among the
hypersurface integrations and auxiliary tasks. For the assessment of
the accuracy of the results, evolutions using a second grid, with half
the resolution (i.e.,$(1000 \times 45)$) in both directions, will be
used throughout.

The numerical evolution is formulated in an explicit manner, and is
hence subject to a CFL stability condition. In the black hole case
under consideration the origin of the coordinate system is not the
vertex of a light-cone, and the constraint imposed by the CFL
condition is weaker. As usually the case in explicit methods, it
becomes linear in the grid spacing instead of quadratic condition
found in~\cite{JMP}.  The computational cost and accuracy of the
simulations, is comparable to axisymmetric (2+1) calculations based on
linearised theory~\cite{krivan}.

\section{Non-linear evolution results}
\label{results}

It is appropriate to start with the venerable quadrupole perturbation
($l=2$ in equation~(\ref{indata})), and fix $\sigma=0.5$,
$r_c=3.0$. This introduces a finite perturbation sitting squarely on
the potential barrier of classic perturbation theory. The various runs
(the complete list is given in Table I), are parametrised by the
amplitude of the initial data, $\lambda$. The range of amplitudes
explored is defined on the one end by the requirement that non-linear
effects are stronger than numerical truncation error and on the other
by the necessity that the evolution maintains a smooth, well behaved,
ingoing light-cone foliation. For sufficiently strong data the
geometry is seen to develop kinks, along specific angles, starting
first at the innermost radial points (one can always hasten the death
of the foliation by evolving deeper inside the horizon). For
evolutions just below the breakdown amplitude, those features require
increasingly larger angular resolution to resolve. The strong field
effects responsible for the caustic formation at the inner edges of
the domain have not been found to be reflected on the actual {\em
observed} non-linear behaviour. For this reason the results presented
here do not attempt to capture the precise limits of the caustic-free
regime.

\begin{figure}[t]
\vspace{0.5cm}
\centerline{\epsfig{figure=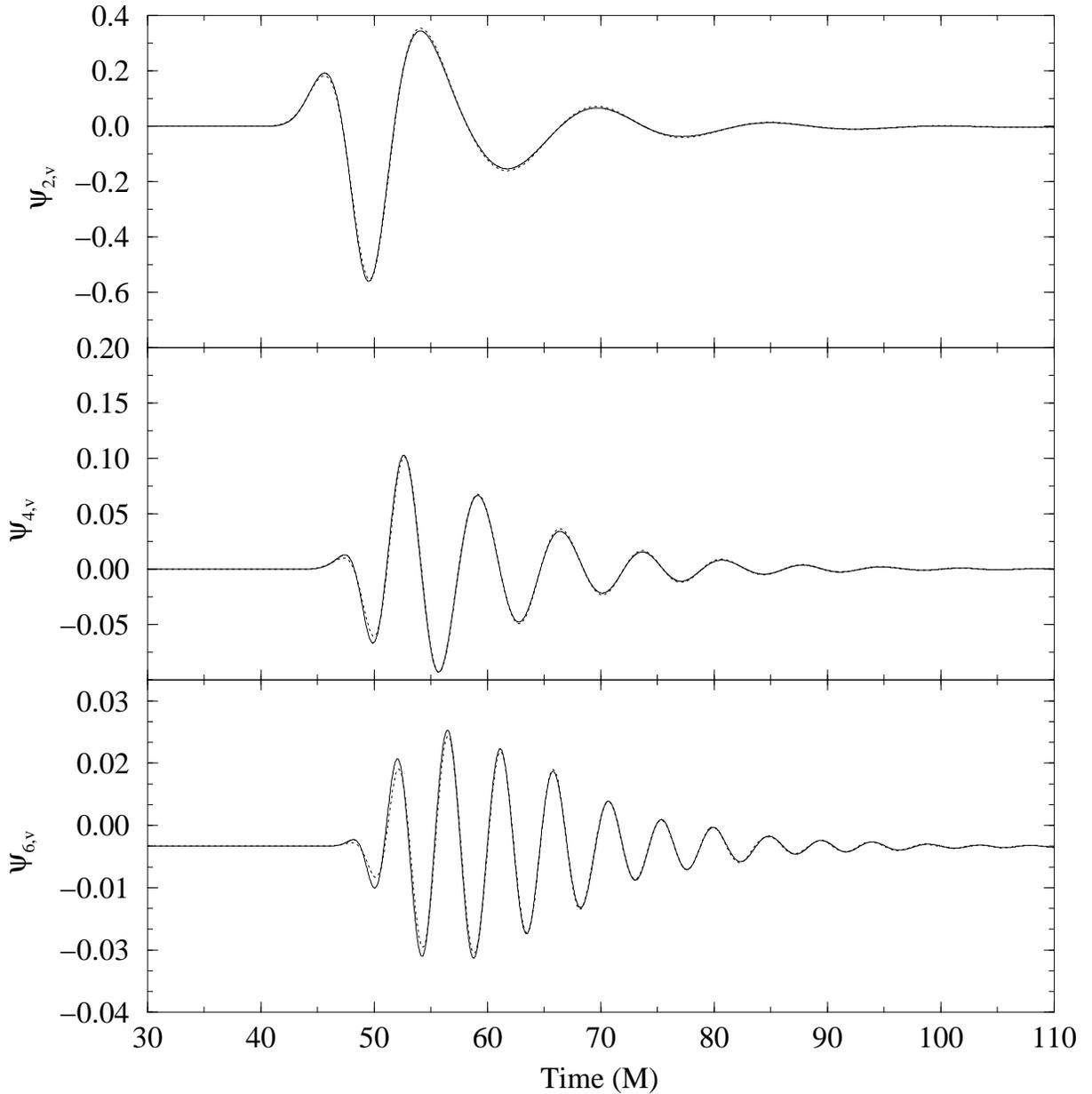,height=17cm,width=17cm}}
\caption{Gravitational radiation signals from a black hole perturbed
with quadrupole initial data. The three different panels show the
time-dependent amplitudes of the harmonically decomposed signal at the
observer (harmonics $l=2,4$ and 6). The upper panel shows the
strongest response (primary harmonic). The tail of all modes exhibits
the typical damped oscillatory behaviour known from linearised
analysis. There are large amplitude differences between the various
modes which are quantified in detail. The higher harmonics exhibit a
{\em time lag} of the first peak with respect to the primary
excitation.  Note also the successive {\em phase reversal} in the
secondary harmonics.  The primary signal has a negative global maximum
which becomes a positive one for $\psi_{4,v}$ and again a negative one
for $\psi_{6,v}$. For the purpose of illustrating visually the code
accuracy, each panel shows two lines, a solid one corresponding to the
high-resolution grid and a dotted one corresponding to half that
resolution.  }
\label{L2response}
\end{figure}

Illustrated here is the strongest case, $\lambda=0.25$. The response
of the black hole to this perturbation is shown in
Fig.~\ref{L2response}, as registered by an observer located at
$20M$. The arrival time of the primary response is roughly twice the
radial separation of the observer from the black hole (40M),
reflecting the fact that outgoing light waves propagate at coordinate
speed 1/2 in an ingoing light-cone framework. The primary (quadrupole)
response carries, for this amplitude, almost 10\% of the black hole
mass in radiation.  An $l=2$ perturbation carries no linear momentum
(it is reflection symmetric with respect to the equator) and hence one
does not expect to see an output of odd harmonics. Given the large
amount of radiated energy, one would expect that the couplings in the
RHS of the evolution equation~(\ref{evolve}) will generate {\em some}
even-harmonic signal. In this case, given that the primary is the
lowest allowed harmonic, all secondary response will have higher $l$
values (up-scattering in $l$ space) and will therefore tend to create
further {\em ripples} in the predominantly spherical black hole
geometry. A first taste of quantitative result follows from the
observed amplitudes of the secondary harmonics ($l=4,6$): They are
respectively one and two orders of magnitude lower in amplitude. Odd
harmonics do not appear, up to roundoff error.

The typical damped oscillatory behaviour known from linearised
analysis is visible here also for large finite perturbations. Looking
closer into the secondaries one notes that, with respect to the
primary wave, the {\em global maxima} of the secondary harmonics i)
are slightly delayed and ii) have phase reversals (i.e., the global
maxima alternate in sign). Given that the primary wave has a peak at
negative amplitude, the origin of the $l=4$ secondary waves in {\em
quadratic} coupling may explain the phase reversal.  Similarly, the
cubic coupling for the $l=6$ wave may account for the second
reversal seen there. This claim will be strengthened by the analysis
of the functional dependence on the initial data amplitude later
on. The evident time delay is harder to be accounted for.

\begin{figure}[t]
\vspace{0.5cm}
\centerline{\epsfig{figure=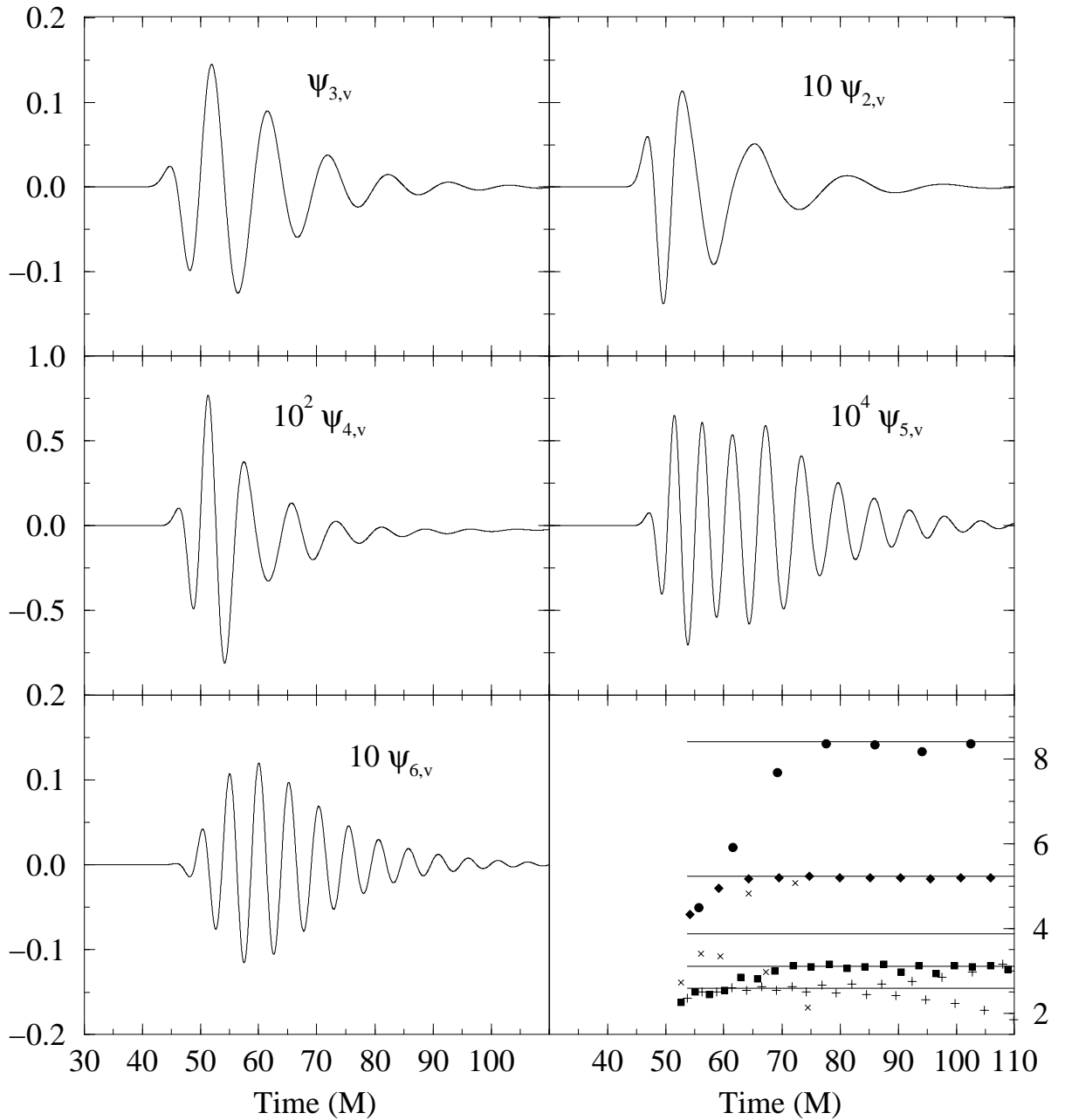,height=18cm,width=18cm}}
\caption{ Gravitational radiation signals from an $l=3$ distortion of
a black hole.  Five different panels show the time-dependent
amplitudes of the harmonically decomposed signal (harmonics $l=2$
through 6). The panel at the upper-left corner shows the primary
harmonic. One notes immediately that for a primary mode with intrinsic
linear momentum (odd $l$ values), {\em all} secondary modes are
excited.  Again the amplitudes of the various harmonics vary
considerably (note the multiplication factors denoted in the panels).
The overall damped oscillation picture is still recognisable but
strong modulations are visible even with cursory inspection. The sixth
panel (lower-right corner) illustrates a time domain analysis of the
frequency content of the signals in the following fashion: Illustrated
are the zero crossings of the various harmonics as a function of
arrival time. Filled circles, diamonds and squares correspond to
$l=2,3,5$ respectively. Crosses denote $l=4$ and Pluses denote
$l=6$. The horizontal lines denote predicted theoretical values for
half-periods from linearised even parity perturbations.  There is
generally good agreement, after about 30M, but substantial deviations
initially. The $l=4$ case exhibits some peculiarity which is discussed
in the text.}
\label{L3response}
\end{figure}

Proceed next to a family of runs where the initial data are set to an
$l=3$ harmonic. This case differs markedly in physical content from
the one described previously, as the initial data do not exhibit
equatorial reflection symmetry and posses net linear momentum. In
addition, the fact that the $l=3$ harmonic is not the lowest allowed
oscillation mode raises the possibility that the secondary harmonics
have both higher and lower angular structure. The down-scattering to
lower $l$ values is particularly interesting, as it represents a
mechanism by which energy is lost from the primary harmonic to a {\em
smoother} configuration. The response, for a family of data of
identical radial profile is monitored again, at the same finite
radius. The five different profiles shown in Fig.~\ref{L3response}
are the time-dependent amplitudes of the harmonically decomposed
signal at the observer (harmonics $l=2$ to 6) for an amplitude of
$\lambda=0.25$. The strongest signal corresponds to the multipole
geometry of the initial perturbation. The present situation is in
contrast with the previous example, where the equatorial reflection
symmetry (zero linear momentum) of the primary mode was reflected in
all secondaries. Despite the fact that the primary mode carries linear
momentum, the secondaries are anything but reflection
anti-symmetric. In fact, the nearest antisymmetric mode ($l=5$) is the
weakest in amplitude. Whereas the overall damped oscillation theme is
still dominant, some subtleties are more apparent here. This can be
seen most easily in the large change in the periods of the $l=2$ mode
between the former and latter parts of the signal and also in the
significant modulation seen in the $l=5$.

It is not entirely clear how to effectively characterise damped
oscillating signals with brief duration and strong variability.  The
sixth panel in Fig.~\ref{L3response} attempts an, at least,
qualitative analysis in the following manner.  For each harmonic, the
zero crossing of the signal at the observer is recorded. Successive
zero crossings provide an instantaneous value for the half-period of
the oscillation. This is plotted as a function of time, illustrating
in this manner the {\em evolution} of the signal frequency in
time. The horizontal lines in that panel are half-periods for QNM
oscillations of even parity perturbations of Schwartzschild black
holes~\cite{Chandra}. The QNM periods are
$T=16.81,10.48,7.76,6.21,5.18$ (in units of M) for perturbations of
$l=2,\cdots,6$ type, respectively. The very strong (doubling) of the
frequency of the $l=2$ harmonic is notable.  The frequency asymptotes
to the weak field value, but only after substantial evolution
time. The primary harmonic ($l=3$) is less strongly evolving and
attains its weak field value earlier on. The $l=5$ response also
exhibits the strong time evolution seen in the quadrupole and reaches
its weak field value late in the evolution. The $l=4$ periods are
problematic and only the few initial estimates are shown. This is
despite the, at first sight, normal appearance of the signal. The
reason has been traced to the secular evolution of the black hole
remnant (region IIb of Fig.~\ref{Diagram}) which affects the time
derivative of the field at the observer. Recall that the effective
energy content between the world-tube and the horizon acts tidally on
the black hole and distorts it. There is a small physical amount of
energy loss, as radiation is absorbed by the horizon, and possibly to
numerical dissipation. Both effects would lead to a slow evolution of
the remnant, which is captured by the time dependent value of
$\gamma_{,v}$ at the observer. The effect will be more pronounced for
harmonics that match closer to the geometry of the remnant (here the
$l=4$) and are weak in amplitude. The tail of the $l=4$ signal is seen
not to asymptote to zero, but rather to a small negative
value. Examining the second time derivative and/or using further
placed observers may help reduce this effect, but frequency
considerations are not central to this study and this is not pursued
further. An oscillatory effect of possibly the same or similar origin
affects the late time periods of the ($l=6$) harmonic which are seen
to oscillate about the QNM value.

Keeping with the same overall setup, the evolution of an $l=4$ primary
is studied. This case is physically close to the quadrupole one,
except that here it is possible to have down-scattering to the
quadrupole harmonic. This adds new scope to the analysis, as it allows
comparison of the up-scattering mechanism from 2 to 4, to the
down-scattering from 4 to 2. One relevant difference of the $l=4$
primary is that caustics appear to be forming here for weaker data
(about a tenth the total energy of the $l=2,3$ cases). This is
possibly due to more efficient focusing of ingoing rays by data of
same total energy if those data have angular structure.

The cumulative energy calculations for the three scenarios are
presented in Table I. This table is the central result of this paper
and hence warrants some discussion. The range of chosen amplitudes has
already been justified. The range of harmonics studied is a compromise
between reasonable completeness and finite resources.  The error bars
are important in establishing confidence in those particular
numbers. They are evaluated by performing the same calculation again,
with exactly twice the resolution and taking the difference in
calculated energy as a measure of confidence. Numerical effects can
artificially generate or suppress harmonics, e.g., through boundary
effects and dispersion. In addition, dissipation will affect the
amplitudes. Low resolution is particularly dangerous for
down-scattered modes, and it is found that there is a minimum angular
resolution required for the non-linear signal to rise above the
truncation error. The error bars serve as a guard against attempting
to resolve too subtle an effect.

\begin{figure}[t]
\vspace{0.5cm}
\centerline{\epsfig{figure=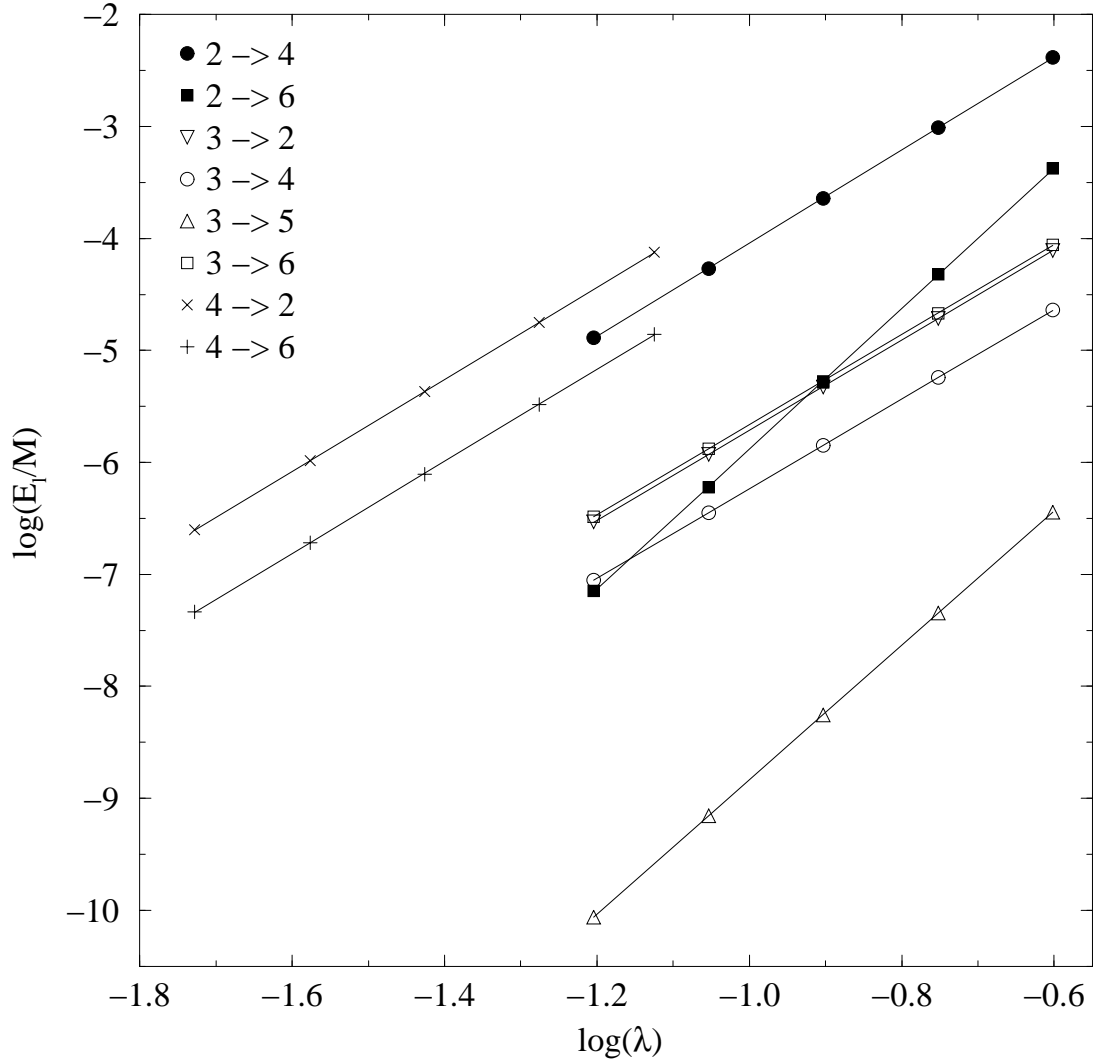,height=17cm,width=17cm}}
\caption{Energy emitted in secondary harmonics as a fraction of the
black hole mass, versus the amplitude of the initial data.  The
straight lines are linear best fits, slopes and intercepts are given
in Table II.  All data points fit well with power laws, reflecting
polynomial underlying couplings. The amplitude range of the $l=4$
family terminates at lower value than the $l=2,3$ families because of
higher sensitivity to caustic formation.  One of the most remarkable
facts uncovered in this non-linear coupling analysis is that
apparently the down-scattering of $l=4$ to $l=2$ is the most efficient
process among those studied.}
\label{power}
\end{figure}

For the range of data studied, the energy in any mode is always a
monotonically increasing function of the initial data amplitude. There
is energy in all modes, except for those couplings excluded by the
conservation of linear momentum. The forbidden couplings typically
show energy well below $10^{-20}$, which suggests an origin in
roundoff error. Indirectly this suggests that the discretized version
of the equations (and its implementation) respects a conservation law
as well. Almost the entire content of Table I can be grasped easily by
putting the data on a log-log graph as in Fig.~\ref{power}. This graph
excludes the primary harmonic data (diagonal elements of the table),
as those encode how much energy is {\em subtracted} from the mode as a
function of initial data amplitude, whereas the off-diagonal terms
capture how much energy is {\em added} to the mode. It is immediately
seen that the energy versus amplitude relation is a power law for all
cases, but with different slopes and normalisations that can be
described by
\begin{equation}
E_{ll'}  =  \epsilon_{ll'} \lambda^{s_{ll'}} M \, ,
\end{equation}
where the first (second) indices denote the primary (secondary)
harmonics respectively. The exponent $s_{ll'}$ will, in general,
follow from the type of coupling, i.e., it reflects a geometric
feature of the theory. The normalisation factor $\epsilon_{ll'}$
captures the efficiency of the coupling and will be much closer
associated with the detailed structure of the solution, in the region
where the coupling is active. With linear best fits to the data
points, one can estimate values for $(s_{ll'},\epsilon_{ll'})$ (Table
II).  

Upon examining the table of numerical fits it is clear that the
quadratic couplings dominate the various transitions (a quadratic
coupling generates a fourth order energy scaling). There are two
examples of higher couplings (cubic in the amplitude), namely the
$2\rightarrow6$ and the $3\rightarrow5$. There are several important
points about efficiencies: {\em The most efficient process is a
down-scattering, namely the $4\rightarrow2$} coupling. This means, in
particular, that for an equal amplitude input, there is more
non-linear energy loss from an octupole to a quadrupole than the
reverse, by a factor of more than two. There is near parity between
the $3\rightarrow2$ and $3\rightarrow6$ couplings, i.e., this shows
that the non-linear energy drain from the momentum-carrying $l=3$
mode is almost equally split between the zero-momentum $l=2$ and $6$
modes.


\section{Summary and Discussion}
\label{summary}

A computational framework for the study of finite amplitude black hole
perturbations has been developed on the basis of the characteristic
initial value problem for the Einstein equations in an ingoing
light-cone setup. A stable code with controllable resolution
requirements was configured for delivering better than 1\% signal
accuracy for angular structures up to $l=6$. Finite amplitude
perturbations were studied for total radiative energies up to 10\% of
the black hole mass. The examination of the distribution of energy in
non-linearly generated radiative modes provides some new insights on
the dynamics of strongly perturbed black holes.

The main conclusions are that: i)The energy transfered through various
couplings is found to obey simple power law scalings, up to the
amplitudes studied. ii) The non-linear couplings are inefficient in
channelling energy away from the primary mode. iii) Up-scattering in
$l$ space (tendency to form ripples in the geometry) and
down-scattering (tendency to smooth the geometry) are found to have
comparable efficiencies, with the distinct possibility that non-linear
smoothing is actually dominant. iv) Various non-linear effects are
present during roughly the first 30M of evolution, including relative
time delays of the different harmonics, phase reversals, and strong
frequency evolution of oscillations before they reach their asymptotic
QNM value. The early periods are always {\em smaller} than the final
QNM value, compatible with a picture in which the initial black hole
is smaller.

A relevant question is whether one can reasonably extrapolate to an
arbitrary amplitude regime. The studies of black hole {\em head-on
collisions} are probing a different, possibly stronger, regime of
distortions (a proper assessment would require a study of invariants
of the horizon geometry). It seems that the phenomenology of the
response in such simulations is compatible with the picture laid out
here. I.e., collisions are genuinely non-linear evolutions, which fail
to generate visibly non-linear signals because of inefficient
off-diagonal couplings. The fairly independent evolution of individual
harmonic modes, {\em a true non-linear property of the Einstein
equations as applied to the black hole system}, clarifies the larger
than expected domain of applicability of infinitesimal black hole
perturbations to collisions.

One can sharpen the picture by discussing a hypothetical non-linear
black hole regime that, apparently, does not happen. Motivated by
other non-linear hyperbolic systems, one could be excused to imagine
an efficient and unhindered energy cascade towards large $l$. The
geometry of a sufficiently strong initial quadrupole perturbation
could say ``pinch'' along certain angles, in {\em a region visible to
the exterior}, hence emitting large amounts of energy in higher
frequencies. This does not happen. Up-scattering efficiencies, besides
failing to dent the energy carried by the primary mode, seem to be
also kept in check by the sligthly more efficient down-scattering
processes. The conjecture is that this ``diagonally dominant''
non-linear transfer matrix, with its appropriately tuned weights above
and below the diagonal, is ultimately responsible for enforcing the
non-linear stability of a black hole to arbitrary distortions.

\acknowledgements 

The author acknowledges support from the European Union via the
network ``Sources of gravitational radiation'' (HPRN-CT-2000-00137)
and a ``Newly Appointed Lecturer'' award from the Nuffield Foundation
(NAL/00405/G). He thanks J.Vickers and N.Anderson for
discussions. Computations were performed on a Compaq XP1000
workstation at the University of Portsmouth.


\newpage

\begin{table}
\begin{tabular}{cc|c|c|c|c|c}
\multicolumn{2}{c|}{\bf Input Data}  & \multicolumn{5}{c}{\bf Output Energy in Harmonics (in units of M) }
\\ \hline
 $l$ &  $\lambda$ & $E_{2}$ & $E_{3}$ & $E_{4}$ & $E_{5}$ & $E_{6}$ \\ \hline
     & 0.25   & $(0.97\pm0.01)10^{-1}$ &  0  & $(0.41\pm0.01)10^{-2}$ & 0 & $(0.42\pm0.04)10^{-3}$ \\
     & 0.177  & $(0.45\pm0.01)10^{-1}$ &  0  & $(0.98\pm0.01)10^{-3}$ & 0 & $(0.47\pm0.02)10^{-4}$ \\
  2  & 0.125  & $(0.21\pm0.01)10^{-1}$ &  0  & $(0.23\pm0.01)10^{-3}$ & 0 & $(0.52\pm0.01)10^{-5}$ \\
     & 0.0884 & $(0.10\pm0.01)10^{-1}$ &  0  & $(0.54\pm0.01)10^{-4}$ & 0 & $(0.60\pm0.01)10^{-6}$ \\
     & 0.0625 & $(0.51\pm0.01)10^{-2}$ &  0  & $(0.13\pm0.01)10^{-4}$ & 0 & $(0.71\pm0.01)10^{-7}$ \\ \hline
     & 0.25   & $(0.87\pm0.05)10^{-4}$ & $(0.11\pm0.01)10^{-1}$ 
              & $(0.23\pm0.01)10^{-4}$ & $(0.36\pm0.01)10^{-6}$ & $(0.87\pm0.02)10^{-4}$ \\ 
     & 0.177  & $(0.19\pm0.01)10^{-4}$ & $(0.54\pm0.01)10^{-2}$ 
              & $(0.57\pm0.03)10^{-5}$ & $(0.45\pm0.01)10^{-7}$ & $(0.22\pm0.01)10^{-4}$ \\ 
  3  & 0.125  & $(0.47\pm0.03)10^{-5}$ & $(0.27\pm0.01)10^{-2}$ 
              & $(0.14\pm0.01)10^{-5}$ & $(0.55\pm0.01)10^{-8}$ & $(0.53\pm0.01)10^{-5}$ \\ 
     & 0.0884 & $(0.12\pm0.01)10^{-5}$ & $(0.14\pm0.01)10^{-2}$ 
              & $(0.35\pm0.02)10^{-6}$ & $(0.70\pm0.01)10^{-9}$ & $(0.13\pm0.01)10^{-5}$ \\ 
     & 0.0625 & $(0.29\pm0.02)10^{-6}$ & $(0.67\pm0.01)10^{-3}$ 
              & $(0.89\pm0.05)10^{-7}$ & $(0.86\pm0.03)10^{-10}$ & $(0.33\pm0.01)10^{-6}$ \\ \hline
     & 0.075  & $(0.75\pm0.04)10^{-4}$ &  0  & $(0.86\pm0.01)10^{-2}$ & 0 & $(0.14\pm0.02)10^{-4}$ \\
     & 0.053  & $(0.18\pm0.02)10^{-4}$ &  0  & $(0.42\pm0.01)10^{-2}$ & 0 & $(0.33\pm0.04)10^{-5}$ \\
  4  & 0.0375 & $(0.43\pm0.04)10^{-5}$ &  0  & $(0.20\pm0.01)10^{-2}$ & 0 & $(0.79\pm0.10)10^{-6}$ \\
     & 0.0265 & $(0.10\pm0.01)10^{-5}$ &  0  & $(0.10\pm0.01)10^{-2}$ & 0 & $(0.19\pm0.02)10^{-6}$ \\
     & 0.0187 & $(0.25\pm0.03)10^{-6}$ &  0  & $(0.50\pm0.01)10^{-3}$ & 0 & $(0.46\pm0.06)10^{-7}$ \\
\end{tabular}
\label{data}
\vspace{0.5cm}
\caption{Cumulative data for energy transfer between harmonic modes in
non-linear black hole perturbations. The left column shows the type of
initial distortion (harmonic index and amplitude). The highest
amplitude in each $l$ family corresponds roughly to the strongest data
that will evolve without forming a caustic near the horizon. For $l=2$
and $l=3$ the energy carried by such data is about 10\% of the black
hole mass. For $l=4$ it is an order of magnitude less. The
off-diagonal elements of the table tell the story of secondary
harmonic generation, towards smoothing or rippling the primary. They
have a variety of scalings and efficiencies, which will be captured
better in the graph of Fig.4.  The zero values in the entries denote
roundoff generated measurements that are typically less than
$10^{-20}$. Error estimates are based on results from lower resolution
runs. Error values below 1\% have been rounded up to that value.}
\end{table}

\begin{table}
\begin{tabular}{|c|c|c|} \hline
{\bf Coupling} $ ll'$ & {\bf Exponent} $s_{ll'}$ & {\bf Efficiency} $\epsilon_{ll'}$ \\ \hline
 24   &   4.16    &   1.32\\
 26   &   6.27    &   2.47\\
 32   &   4.02    &   $2.07 \, 10^{-2}$\\
 34   &   4.00    &   $5.49 \, 10^{-3}$\\
 35   &   6.01    &   $1.51 \, 10^{-3}$\\
 36   &   4.02    &   $2.31 \, 10^{-2}$\\
 42   &   4.11    &   $3.11$\\
 46   &   4.11    &   $2.07$\\ \hline
\end{tabular}
\vspace{0.5cm}
\caption{Derived coupling exponents and efficiencies based on the data
of Table I.  Eight couplings are studied in total, listed with a
compact notation in the left column. Six couplings are seen to be
quadratic, the other two being cubic (the energy scales as the square
of the amplitude). The efficiencies range in value. The most efficient
quadratic process (at unit amplitude this would be dominant quadratic
process) transfers energy from the octupole harmonic to the quadrupole
one. Near unit amplitudes, the cubic up-scatterings would appear to
generally dominate. It is highly likely though that the examination of
an $l=6$ primary evolution would reveal even more efficient cubic
down-scatterings.}
\label{fits}
\end{table}

\newpage

\newpage

\newpage

\newpage

\end{document}